\documentclass[12pt]{article}

\usepackage{epsf}
\usepackage[dvips]{graphicx}

\begin{document}

\begin{center}
{\large\bfseries Saturation of Hadron Production\\
in Proton-(anti)Proton Collisions at Low $P_T$}

\vskip 5mm
I. Zborovsk\'{y}$^{1, \dag}$ and M.V. Tokarev$^{2, \flat }$

\vskip 5mm
{\small (1) {\it Nuclear Physics Institute ASCR,
\v{R}e\v{z}, Czech Republic }}
\\
(2) {\it Joint Institute for Nuclear Research,
Dubna, Russia }
\\
$\dag$ {\it E-mail: zborovsky@ujf.cas.cz }
\\
$\flat$ {\it E-mail: tokarev@jinr.ru }
\end{center}

\vskip 5mm
\begin{center}
\begin{minipage}{150mm}
\centerline{\bf Abstract} Experimental data on inclusive cross
sections of the hadrons produced in high energy proton-(anti)proton
collisions are analyzed in the $z$-scaling approach. Saturation of
the scaling function $\psi(z)$ for $z<0.1$ (low transverse momenta)
was found. The first results on charged hadron spectra in $pp$
collisions obtained  by the CMS Collaboration at the LHC have
confirmed the saturation down to the value of $z\simeq 0.05$. The
CMS data on $K^0_s$-meson production at $s^{1/2}=7$ TeV extend the
saturation region even to a lower value of $z\simeq 0.002$ in the
new energy domain. A microscopic scenario of hadron production at a
constituent level based on the $z$-scaling is discussed in the
saturation regime.
\end{minipage}
\end{center}

\vskip 10mm

\section{Introduction}

The inclusive spectra carry information about the particle production mechanism
and provide sensitive tool to probe dynamics of constituent interaction.
Experimental data on hadron distributions from $pp/p\bar p$ collisions
are a benchmark to investigate more complex processes in $AA$ collisions.
One of the methods based on mutual comparison of the data on inclusive cross sections
at different collision energies, multiplicity densities, transverse momenta and detection
angles, is based on  $z$-scaling \cite{Gen_Z}.
The approach reflects self-similarity as one of the basic
symmetries in the hadron production at the constituent level.
The scaling regularity includes the region of high transverse momenta as well
as processes with small momenta and high multiplicities.
The variable
\begin{equation}
z = \frac{  \sqrt {s_{\bot}}} {W}
\label{eq:r2}
\end{equation}
is the ratio of two quantities. The transverse kinetic energy ${\sqrt {s_{\bot}}}$
of the constituent sub-process consumed on production
of the inclusive particle
and its recoil partner (its antiparticle),
is expressed in units of the nucleon mass.
The quantity $W$ is the maximal relative number of the constituent configurations
$\{x_1, x_2, y_a, y_b\}$
which includes the configurations satisfying the kinematical condition
\begin{equation}
(x_1P_1+x_2P_2-p/y_a)^2 = M_X^2.
\label{eq:r4}
\end{equation}
Here $M_X=x_1M_1+x_2M_2+m/y_b$ is a recoil mass,
$P_1$, $P_2$, $p$, and $M_1$, $M_2$, $m$ are the 4-momenta and  masses
of the colliding and inclusive particles, respectively.
The value of $W$ is related to the
entropy of the rest of the system:
\begin{equation}
S=\ln W + \ln W_0.
\label{eq:r15}
\end{equation}
The absolute number of configurations $W_0$ depends on the hadron
type $(F)$ and drops out of the $z$-scaling.
The scaling functions
$\psi(z)$ for different hadrons collapse onto a single
curve by means of the transformation:
$z \rightarrow \alpha_F z$,  \  $ \psi \rightarrow \alpha_F^{-1} \psi $,
where $\alpha_F=W_0^F/W_0^\pi$. The transformation preserves
the energy, angular and multiplicity independence of
$z$-presentation of hadron spectra as well as normalization of $\psi(z)$ to unity.

\section{$z$-Scaling in soft $pp/p\bar p$ interactions}

In this contribution we focus on the soft region of the $pp/p\bar p$ interactions where
collective phenomena at various levels can take place.
The bulk of the produced matter at low-$p_T$ consists of multitude of strongly
interacting constituents. Though there is no direct information on the type of
the constituents, the microscopic scenario based on the $z$-scaling allows us to extract
information on kinematics of the constituent sub-processes.
This is obtained by the assumption on self-similarity of
hadron interactions at the constituent level
and translated into the functional form of variable $z$.

The scaling variable $z$ includes a combination of the kinematical
characteristics of the constituent sub-processes with some
parameters ($c$, $\delta$, $\epsilon_F$) describing the system.
Parameter $c$ has analogy with the specific heat of the produced matter.
The parameters $\delta$ and $\epsilon_F$ are fractal dimensions of
the colliding protons (antiprotons) and the fragmentation process,
respectively. The determination of the parameters by self-similarity
arguments from the measured spectra gives dependencies of the
momentum fractions $\{x_1, x_2, y_a, y_b\}$ on the collision energy
and centrality, transverse momentum, detection angle, and type of
the produced particle. This provides a microscopic scenario of the
underlying constituent sub-processes. This approach is applied for
arbitrary momenta of the inclusive particle. For $pp/p\bar p$
collisions it has been found that $c=0.25$, $\delta=0.5$, and
$\epsilon_F$ depends on the type of the inclusive hadron.
\begin{figure}[h]
\hspace*{.5cm}
\includegraphics[width=70mm,height=65mm]{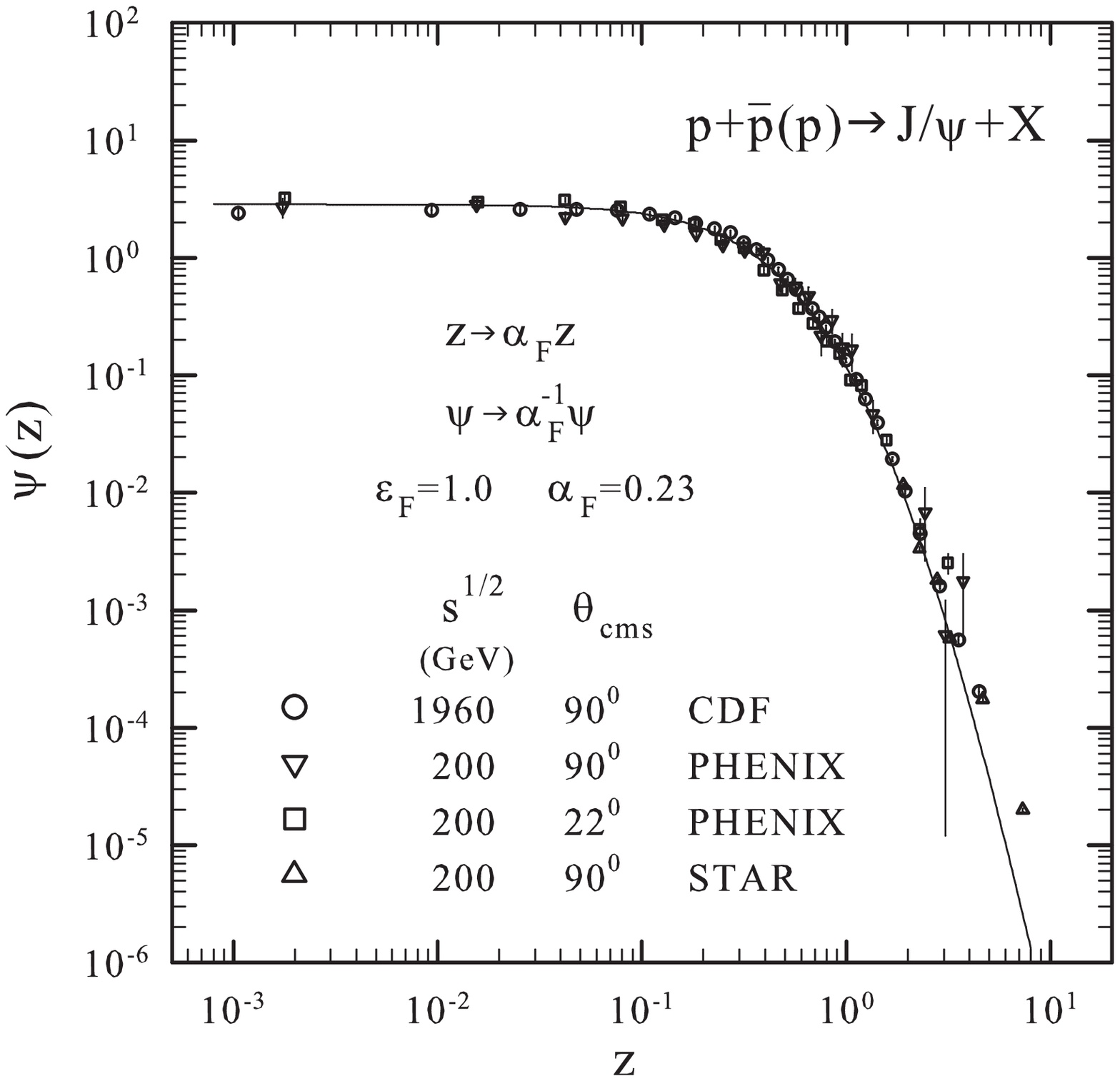}
\hspace*{1.cm}
\includegraphics[width=70mm,height=65mm]{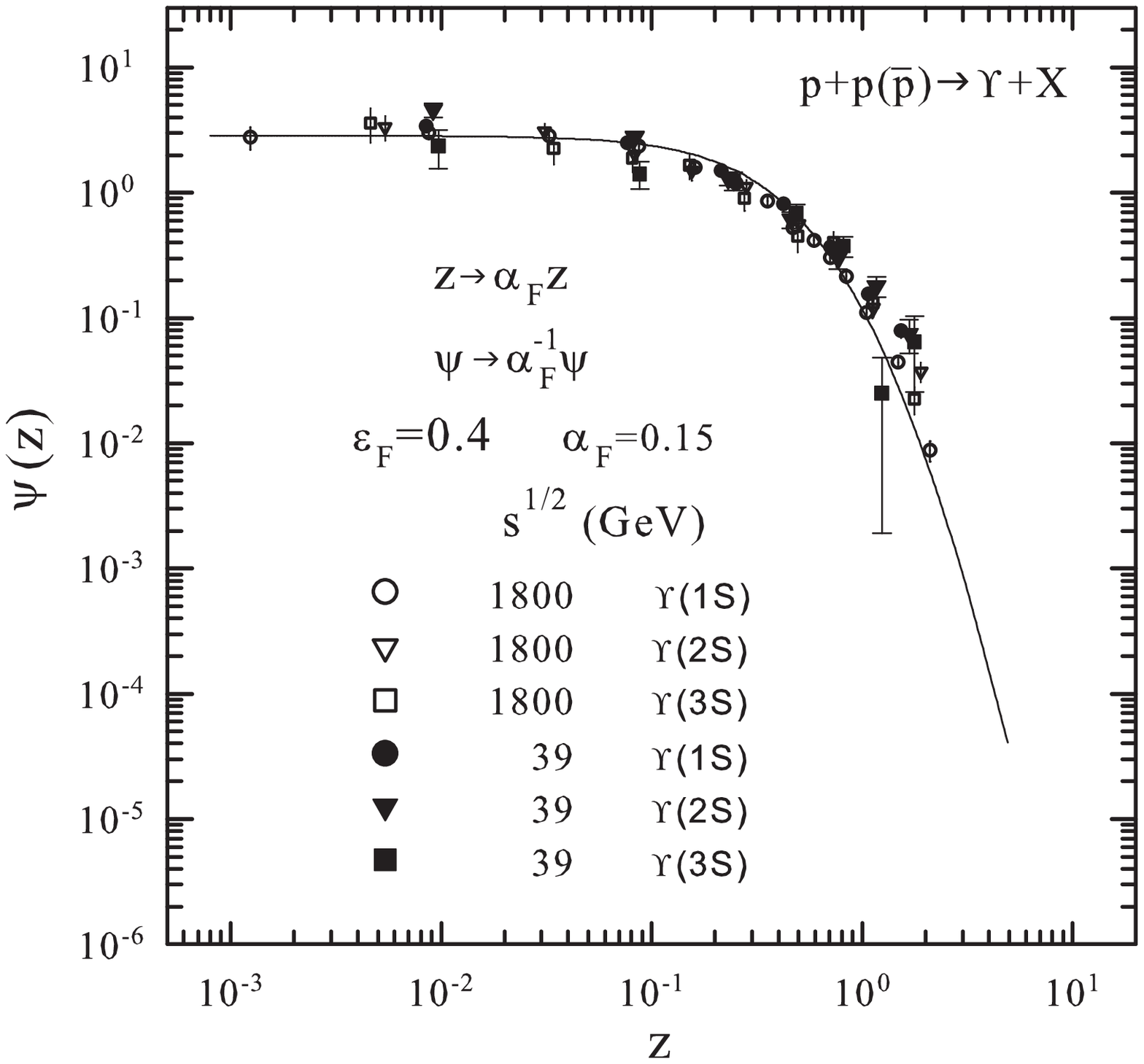}
\vskip -7mm
\hspace{4.5cm}    a) \hspace{7.7cm}   b)
\vskip -2mm
\caption{
(a) The spectra of $J/\psi$-mesons measured
at $\sqrt s = 200$~GeV for $\theta_{cms}=90^0, 22^0$ and
at $\sqrt s = 1960$~GeV for $\theta_{cms}=90^0$ in $z$-presentation.
(b) The spectra of $\Upsilon(1S)$-, $\Upsilon(2S)$-, and
$\Upsilon(3S)$-mesons measured at $\sqrt s = 39$ and  1800~GeV in $z$-presentation.
}
\end{figure}

The soft processes in $pp/p\bar p$ interactions are typical for
the low-$p_T$ particle production with small $z$.
In this region the  saturation of the scaling function $\psi(z)$ for $z<0.1$
was observed \cite{Gen_Z}.
The measurements of spectra for identified particles extend
the approximate constancy of $\psi(z)$ to even lower values of $z$.
The $z$-presentation of inclusive spectra
of pions, kaons, and antiprotons measured at ISR energies
revealed the saturation in the region of $0.01<z<0.1$.
This was confirmed by the measurements of $K^*$ resonance
at RHIC at the value of $z \simeq 0.007$ \cite{K_STAR}.
The inclusive spectra of heavier hadrons ($J/\psi, D^0,  B,
\Upsilon$) obtained at the Tevatron energies
$\sqrt s=$1800 and 1960~GeV have manifested the saturation of the
$z$-scaling in $p\bar p$ collisions down to $z \simeq 0.001$.
\begin{figure}[h]
\hspace*{0.5cm}
\includegraphics[width=70mm,height=65mm]{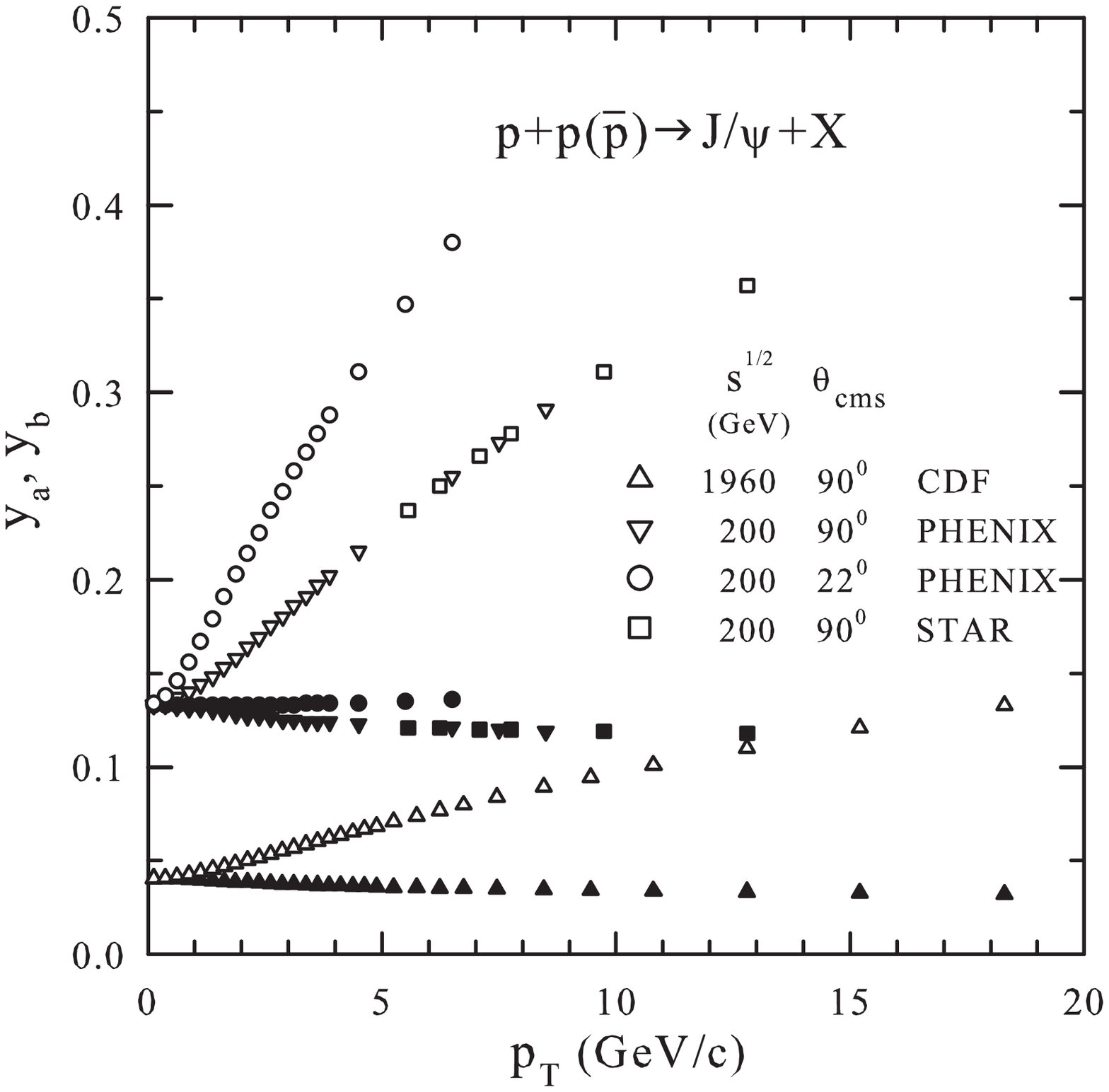}
\hspace*{1.cm}
\includegraphics[width=70mm,height=65mm]{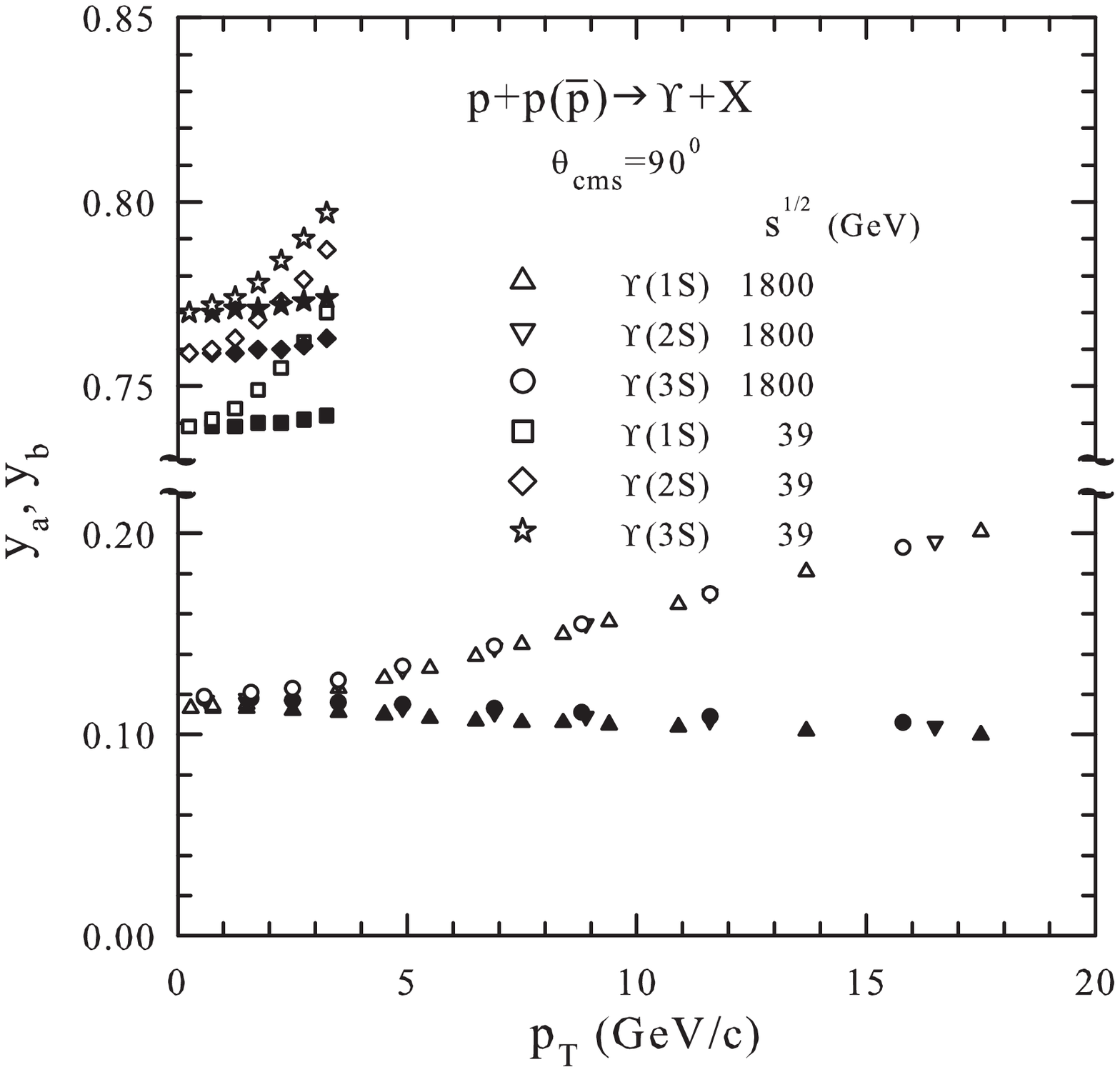}
\vskip -7mm
\hspace{4.5cm}    a) \hspace{7.7cm}   b)
\vskip -2mm
\caption{
The $p_T$-dependence of the momentum fractions $y_a$ and $y_b$ for (a)
$J/\psi$- and (b) $\Upsilon$-mesons produced in $pp/p\bar p$ collisions
by different kinematical conditions.
The empty and full symbols correspond to $y_a$ and $y_b$, respectively.
}
\end{figure}

Figure 1(a) demonstrates results of the combined analysis of the RHIC \cite{STAR_J,PHENIX_J}
and Tevatron \cite{CDF_J} data
on $J/\psi$-meson spectra  measured in $pp$ and $p\bar p$ collisions
at the energies   $\sqrt s =$ 200, 1960~GeV
and  angles $\theta_{cms}=22^0, 90^0$ in $z$-presentation.
Similar results are shown in Fig. 1(b) for different mass states $(1S, 2S, 3S)$
of $\Upsilon$-mesons produced  at $\sqrt s = 39$~GeV \cite{FNAL_Y} and
$\sqrt s = 1800$~GeV \cite{CDF_Y}, respectively.
The solid line in Fig.1(b) is the same fit as shown in Fig. 1(a).
As seen from Figs. 1(a) and 1(b), the $z$-presentation of
the  $J/\psi$ and $\Upsilon$ spectra  manifests saturation
in the range $z=10^{-3}-10^{-1}$ for different collision energies,
production angles, and respective mass states of $\Upsilon$-meson.

The dependencies of the fractions $y_a$ (empty
symbols) and $y_b$ (full symbols) on the transverse momentum $p_T$
of the inclusive particle are shown in Fig. 2.
The momentum fraction
$y_a$ characterizes the  energy loss ($\Delta E \sim 1-y_a$) by
formation of the inclusive particle. The energy loss increases with
the collision energy $\sqrt s$ and decreases with the transverse
momentum $p_T$. It is considerably larger in the central region
($\theta_{cms}=90^0$) than in the fragmentation ($\theta_{cms}=22^0$) one.
Production of the $J/\psi$-meson is accompanied with extra large
energy losses and recoil mass $M_X$ when compared with other hadrons.
It corresponds to relatively small values of $y_a$, $y_b$
and the large value of $\epsilon_{J/\psi}=1$
as required by the energy independence of $\psi(z)$.
For $\Upsilon$-meson, the energy loss is sensitive to its respective
mass state at $\sqrt s = 39$~GeV.
It was found to be the smallest for the $3S$ state.
The energy loss increases and is equalized for all three mass states
of $\Upsilon$ at $\sqrt s = 1800$~GeV.

The values of $y_b$ become considerably smaller than the values of
$y_a$ as $p_T$ increases. This means that the momentum balance in
the production of the inclusive particle from the sub-process is more
likely compensated by the states with higher multiplicity of particles
having smaller momenta than by a single particle with a higher
momentum moving in the opposite direction. The observed property is
directly related with the recoil mass in the constituent
collision. For high collision energies it is well approximated by
$M_X\simeq m/y_b$. At low transverse momenta both fractions
$y_a$ and $y_b$ become equal to each other. It means that, at low
$p_T$, the objects produced in the constituent collision into the
near- and away-side direction, have equal masses. As a consequence,
the approximate relation $v_p=p/m \simeq v_q=q/M_X$ is valid, where
$q$ is 4-momentum of the fragmenting objects in the scattered or
recoil directions.
 This implies equal velocities $v_p$  and  $v_q$
of the detected particles and their fragmenting ancestors though
the mass of the ancestors $M_X$ increases with the collision energy.
This kinematics is in tune with the ideas of coherence in production
of particles at low $p_T$.

\section{New LHC data and saturation of $\psi(z)$ at low $z$ }

In this section we analyze the first data on
transverse momentum distributions of the charged
hadrons and neutral kaons produced in $pp$ collisions at the LHC.
The CMS Collaboration measured the spectra \cite{CMS_Hpm}
of charged hadrons
at the energies $\sqrt s = 900$ and 2360~GeV in the
central rapidity range.
Figure 3(a) shows $z$-presentation of the spectra in comparison with the
data from RHIC \cite{STAR_Hpm}, ISR \cite{BS_Hpm}, and FNAL (the fixed target) \cite{FNAL1_Hm,FNAL2_Hpm}
in the energy range
$\sqrt {s_{NN}} = 19-2360$~GeV at $\theta_{cms}\simeq 90^0$.
The CMS data have revealed similar tendencies as the data at lower energies.
As it is seen from Fig. 3(a) the first LHC data on the charged hadron distributions
have confirmed the energy independence
of the scaling function with the same values of
parameters $\delta$, $\epsilon_F$, and $c$.
At low $p_T$, the data extend the saturation region of $\psi(z)$
for non-identified hadrons down to a value of $z \simeq 0.05$.
The saturation of the scaling function is examined in the new energy range.
The behavior of $\psi(z)$ at even lower $z$ can be investigated
by increasing the collision energy $\sqrt s $, or the multiplicity
density $dN_{ch}/d\eta$, or by decreasing the transverse momentum.
There is a special opportunity for neutral kaons to be measured
for very small $p_T$.
\begin{figure}[h]
\hspace*{0.5cm}
\includegraphics[width=70mm,height=65mm]{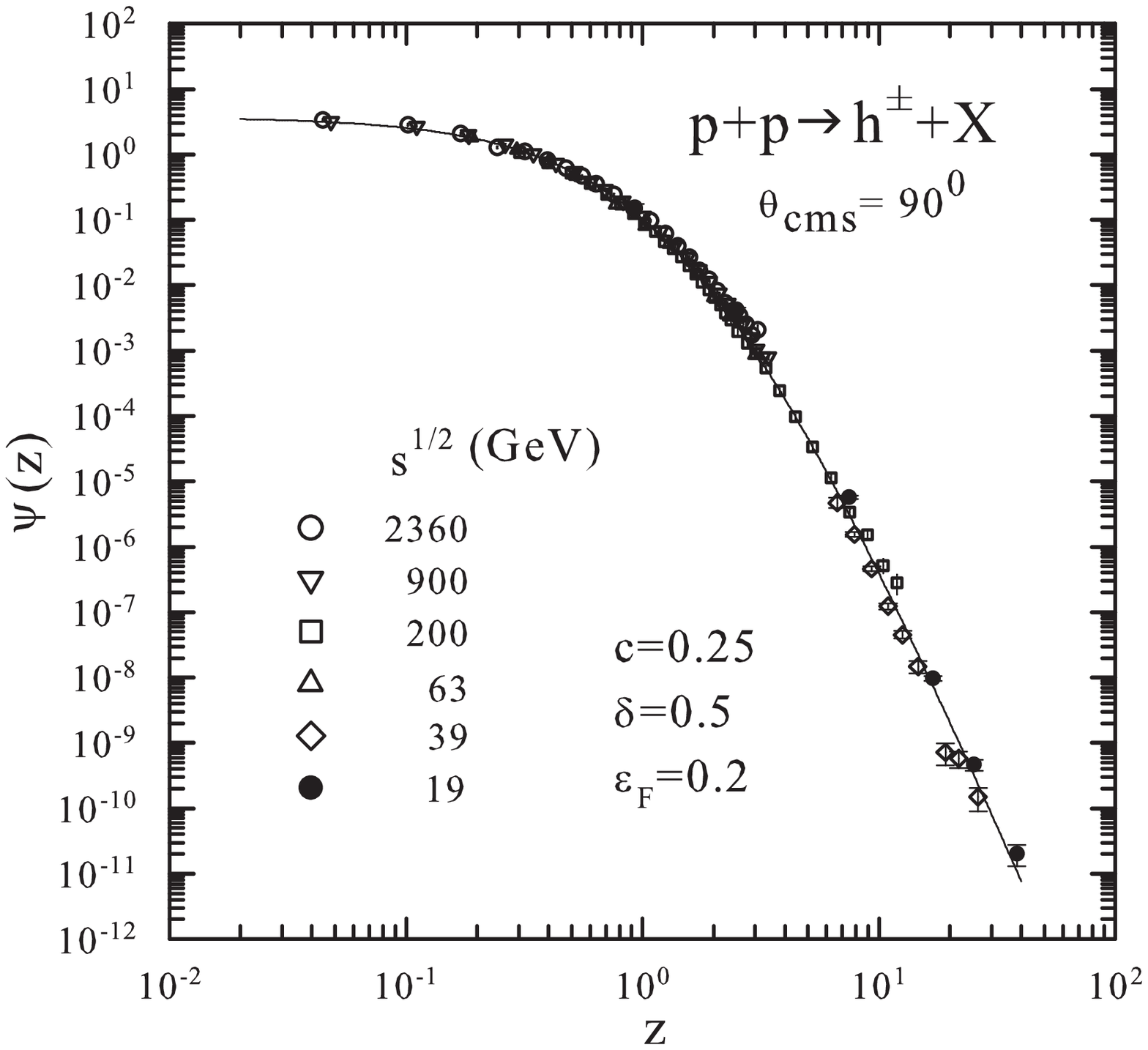}
\hspace*{1.cm}
\includegraphics[width=70mm,height=65mm]{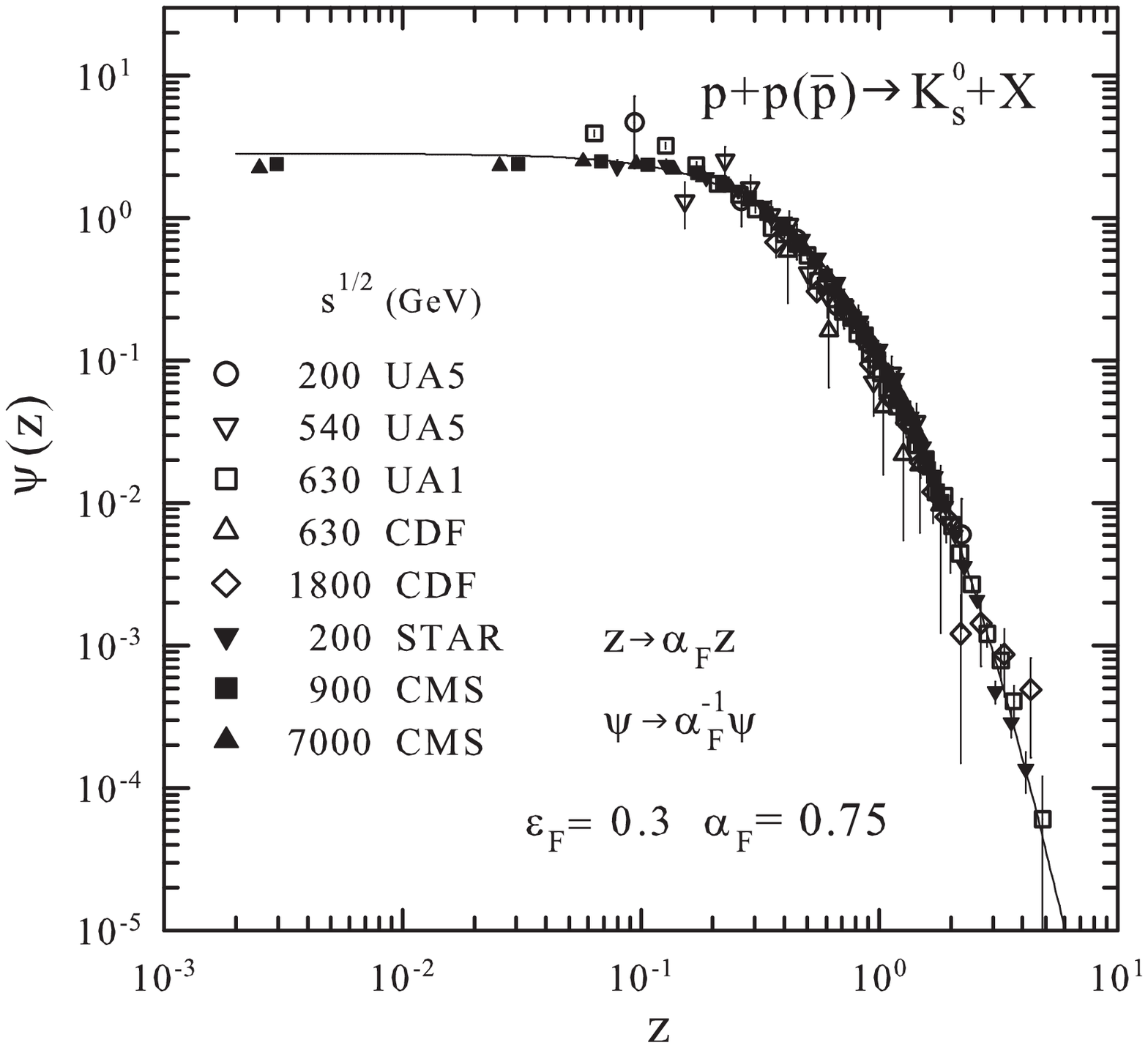}
\vskip -9mm
\hspace{4.5cm}    a) \hspace{7.7cm}   b)
\vskip -2mm
\caption{
(a) The spectra of the charged hadrons produced in $pp$ collisions
at $\sqrt s =19-2360$~GeV and $\theta_{cms}=90^0$
in $z$-presentation.
(b) The spectra of $K^0_s$-mesons produced in $pp/p\bar p$ collisions
at $\sqrt s =200-7000$~GeV and $\theta_{cms}=90^0$
in $z$-presentation.
}
\end{figure}

The CMS Collaboration at LHC measured the spectra of $K^0_s$-mesons \cite{CMS_K0s}
produced in $pp$ collisions at the energies $\sqrt s = 900$ and 7000~GeV
in the central rapidity range.
The data  include measurements at small
transverse momentum $p_T\simeq 50$~MeV/c.
Figure 3(b) shows $z$-presentation of the spectra in comparison with
the data from the Collaborations STAR
\cite{STAR_K0s}, UA5 \cite{UA5_K0s}, UA1 \cite{UA1_K0s},
and CDF \cite{CDF_K0s} in the energy range
$\sqrt {s_{NN}} = 200-7000$~GeV at $\theta_{cms}\simeq 90^0$.
\begin{figure}[h]
\hspace{1.cm}
\includegraphics[width=70mm,height=65mm]{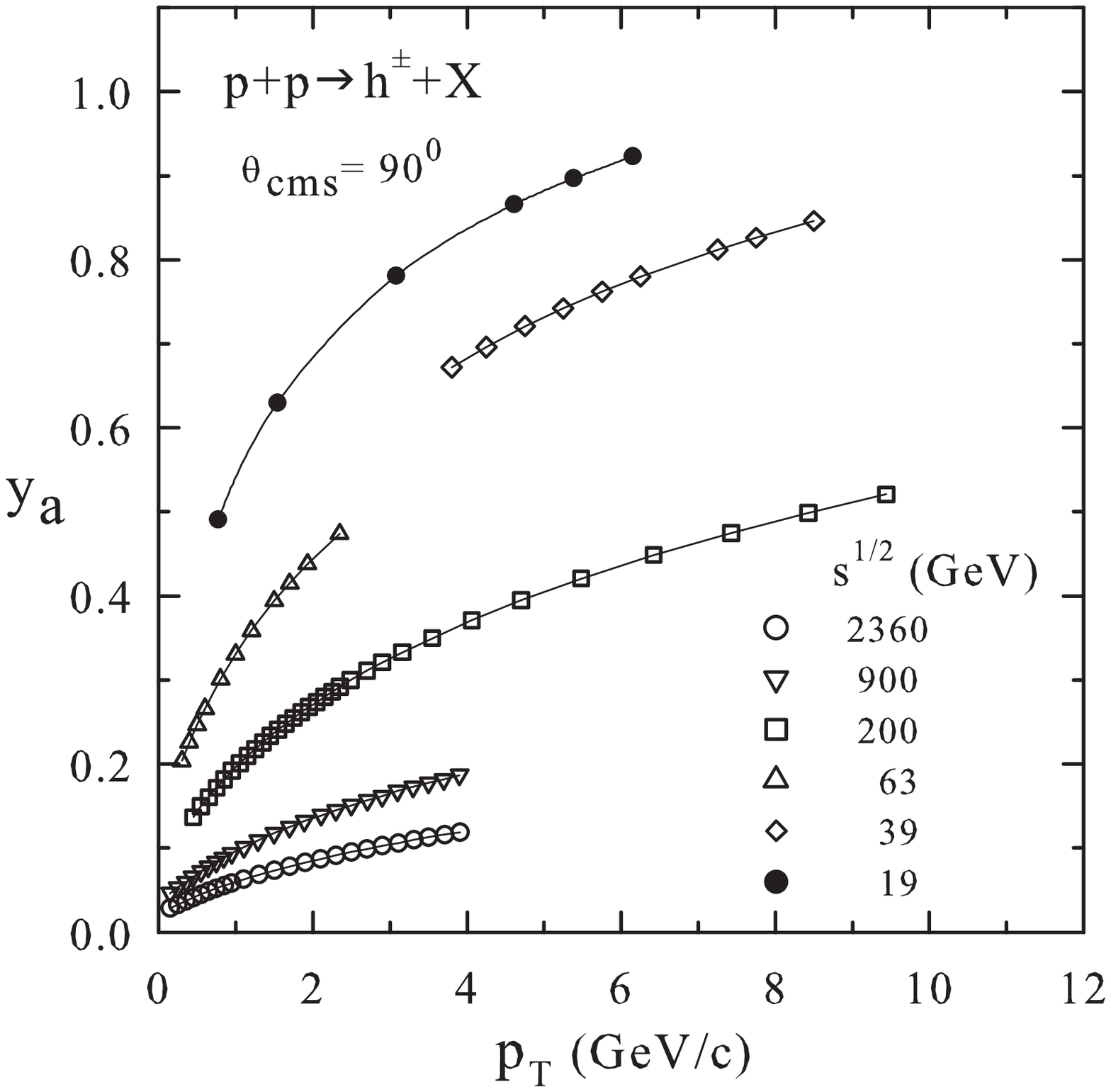}
\hspace*{1.cm}
\includegraphics[width=70mm,height=65mm]{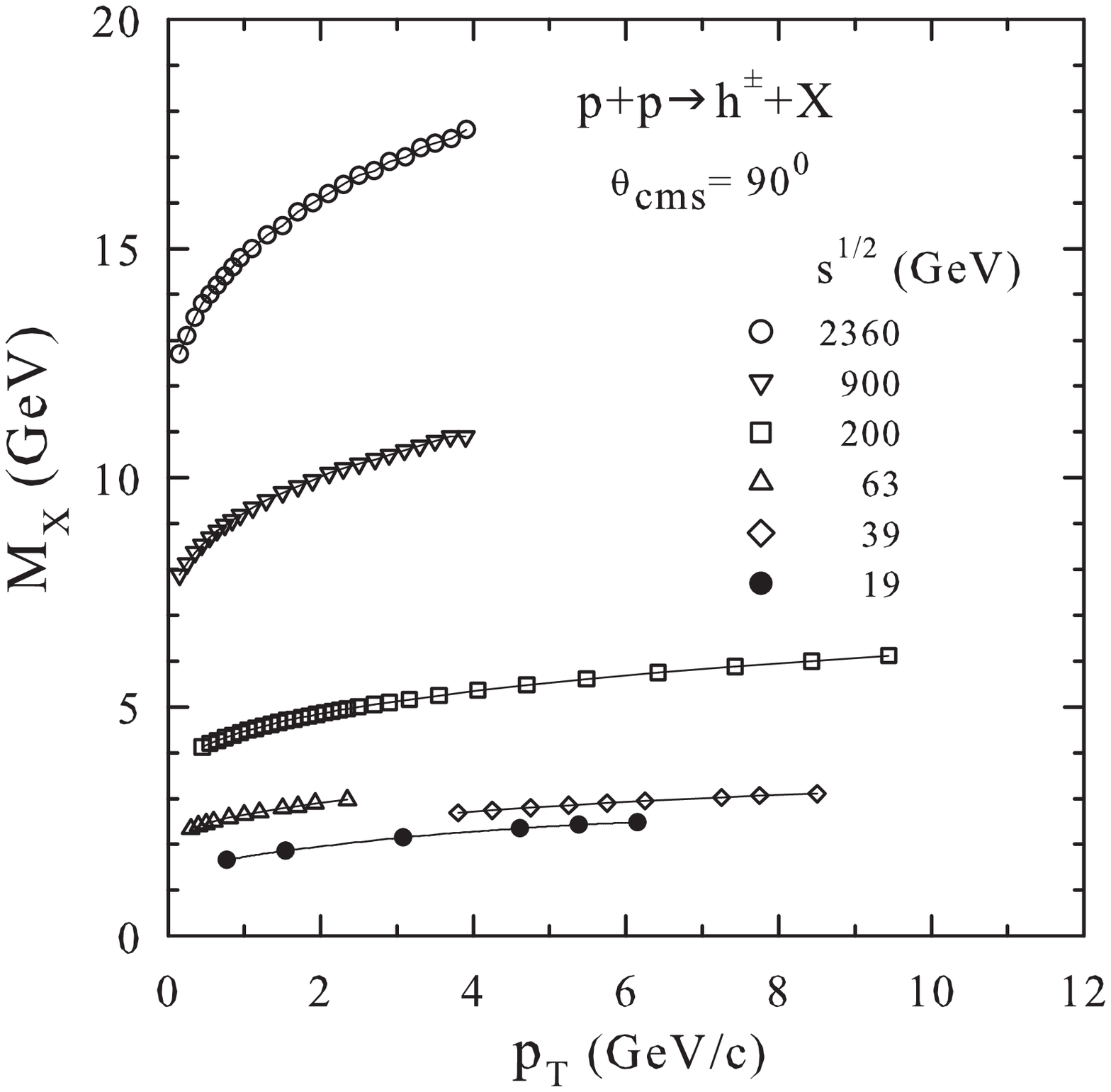}
\vskip -7mm
\hspace{4.5cm}    a) \hspace{7.7cm}   b)
\vskip -2mm
\caption{
The dependence of the fraction $y_a$ (a) and the recoil mass $M_X$ (b)
on the transverse momentum $p_T$
for the charged hadrons produced in the $pp$ collisions
at $\sqrt s = 19 - 2360$~GeV.
}
\end{figure}
In the measured $p_T$ range, the new LHC spectra are consistent
with the $z$-scaling observed at lower energies.
The saturation of the scaling function $\psi(z)$
for $K^0_s$-mesons is confirmed down
to a value of $z\simeq 0.002$.
When compared with Figs. 1(a) and 1(b),
the constancy of $\psi(z)$ is verified in the new LHC energy domain
 for $z$ $\simeq (0.002-0.1)$.
\begin{figure}[h]
\hspace{1.cm}
\includegraphics[width=70mm,height=65mm]{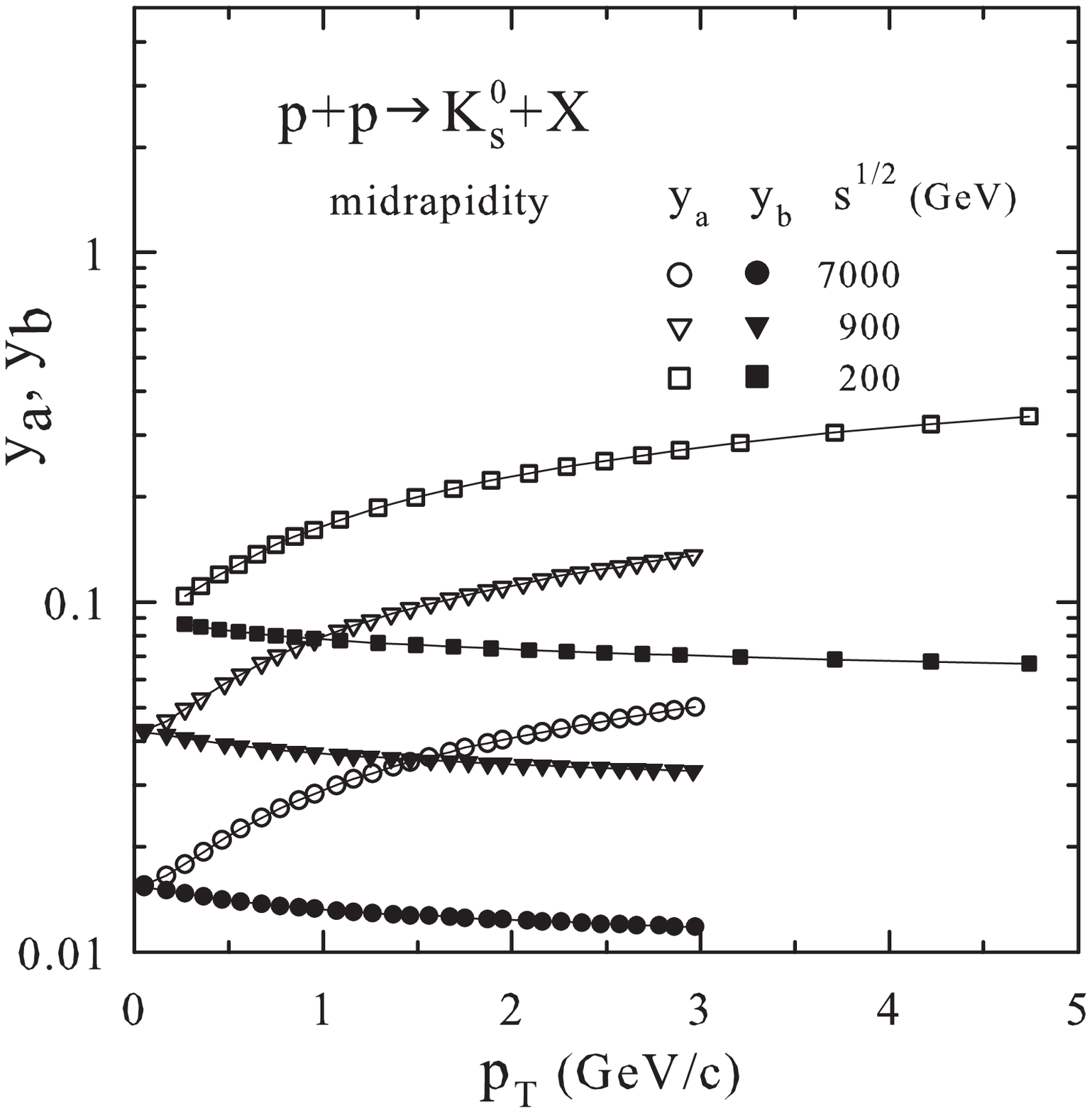}
\hspace*{1.cm}
\includegraphics[width=70mm,height=65mm]{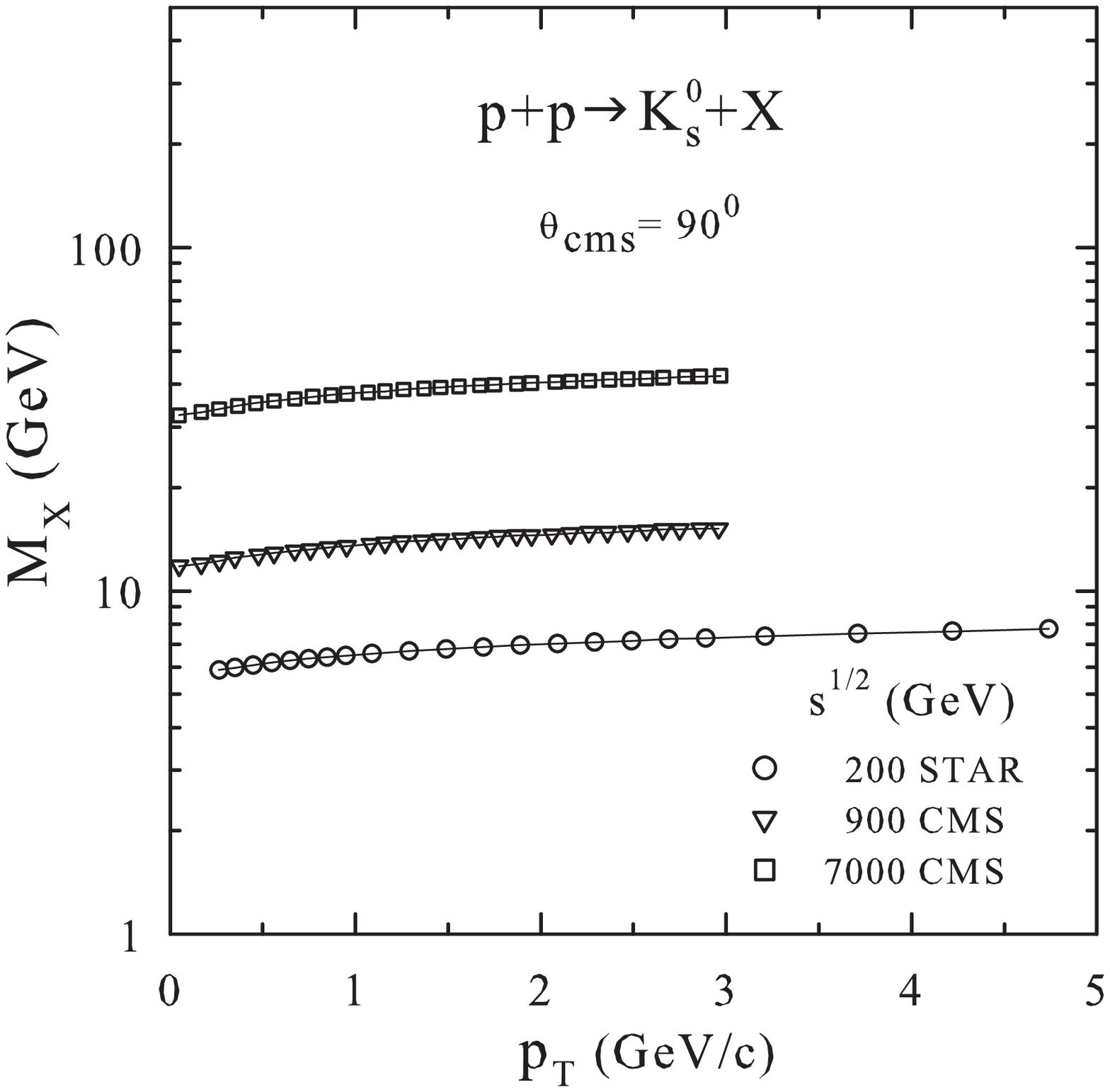}
\vskip -7mm
\hspace{4.5cm}    a) \hspace{7.7cm}   b) \vskip -2mm
\caption{
The $p_T$-dependence of the fractions $y_a$, $y_a$ (a) and the
recoil mass $M_X$ (b) for $K^0_s$-mesons
produced in $pp$ collisions at the energies $\sqrt s = 200, 900$, and
7000~GeV.
}
\end{figure}

Figure 4(a) shows the growth of the energy loss $\Delta E \sim 1-y_a$
with $\sqrt s $.
For $p_T\simeq 4$~GeV/c, the energy loss is about 20\% at $\sqrt s = 19$~GeV and
90\%  at $\sqrt s = 2360$~GeV.
Figure 4(b) demonstrates the $p_T$-dependence of $M_X$
for the charged hadrons produced in $pp$ collisions
at $\sqrt s = 19-2360$~GeV.
The recoil mass at the LHC energy is considerably larger
than at RHIC and SPS energies.
For $p_T\simeq 4$~GeV/c it was found to be about $M_X \simeq 18$~GeV at
$\sqrt s = 2360$~GeV which is much higher than the value of
$M_X \simeq 2$~GeV at $\sqrt s = 19$~GeV.
The similar tendencies are seen for $K^0_s$-mesons from Figs. 5(a) and 5(b).
As at lower energies (Fig.2), equality $y_a\simeq y_b$
and large $M_X$ at low $p_T$
indicate the coherence in the soft processes at $\sqrt s =7000$~GeV.

\section{Conclusions}

New data
on charged hadron and $K^0_s$-meson production
in $pp$ collisions
measured by the CMS Collaboration
at the LHC, have  confirmed the saturation
of the scaling function $\psi(z)$ observed at lower energies
at the ISR, SPS, RHIC, and Tevatron.
A microscopic scenario of the hadron production based on
the $z$-scaling was used to estimate the characteristic
increase of the energy loss and
recoil mass in the constituent interactions in the low-$z$ region
with the increasing collision energy $\sqrt s$ .
The universal scaling behavior in the saturation region suggests
that mechanism of the particle production at low $p_T$ is governed by
soft self-similar processes  which reveal some kind of a mutual equilibrium.
The momentum fractions $y_a$ and $y_b$ at low $p_T$
indicate the coherence in the processes with soft particle production.

\section{Acknowledgments}

The investigations have been
supported by the IRP AVOZ10480505,
by the Ministry of Education, Youth and Sports
of the Czech Republic grant LA08002
and by the special program of the
Ministry of Science and Education of the Russian Federation,
grant RNP.2.1.1.2512.

\end{document}